\documentclass[aps,prf,onecolumn,showpacs,superscriptaddress,groupedaddress,floatfix]{revtex4}  
\usepackage{amsmath}
\usepackage{graphicx}  
\usepackage{dcolumn}   
\usepackage{bm}  
\usepackage{amssymb}   
\usepackage{graphicx}
\usepackage{float}
\usepackage{caption}
\usepackage{lineno}  
\usepackage[dvipsnames]{xcolor}
\usepackage{comment}

\begin{document}  
 

\widetext


\title{Role of Transient Dynamics in Dripping-Jetting Transition in Newtonian Fluids}

\author{Kishorkumar Sarva}
 \affiliation{%
Interdisciplinary Center for Energy Research (I.C.E.R), Indian Institute of Science, Bangalore, India
}%
\author{ Tejas G Murthy}
\affiliation{
Civil Engineering, IISc Bangalore, 560012, India
}
\author{Gaurav Tomar}
 \email{gtom@iisc.ac.in}
 \affiliation{%
 Department of Mechanical Engineering, Indian Institute of Science, Bangalore, India
}%

\begin{abstract}

Dripping dynamics has been well studied over the past century and forms a classic example of chaotic system in physics. With an increase in the inlet flow rate, periodic droplet formation from a faucet becomes chaotic in terms of the droplet size and the length of the liquid column at the time of pinch-off. With a further increase in the flow rate, dripping regime transitions into jetting regime where the liquid column length is much longer than that observed in the dripping case. In general, dripping faucet is seen as a long time behavior of the system at fixed control parameters. In the steady state condition, different nonlinear behaviors such as periodic and chaotic formation of droplets are observed in the dripping and jetting regimes. It is known that dripping faucet shows chaotic dripping regime before jetting regime ensues. 
At a critical inlet velocity, $U_{m-d_j}$, we note that dripping to jetting transition occurs after several droplets have formed in the dripping regime. The transition behaviour can be characterized by the time evolution of the liquid jet length $L$ and droplet size $D_p$. Solution to slender jet equation show that the dripping-jetting transition region is a function of the fluid properties. Further, we show that perturbations in the inlet velocity can significantly modify the transient behavior of the dripping to jetting regime transition.
 
\end{abstract}

\maketitle

\section{Introduction}
 A comprehensive understanding of the mechanism of drop formation from a nozzle has myriad applications in ink-jet printing, additive manufacturing, spray coating, spray cooling, and pesticide spraying. The critical challenge in these processes is accurate control of  size and frequency of droplet formation. The process of droplet formation from a nozzle is called `dripping'. These industrial processes have been the underpinning for fundamental studies on dripping dynamics of a faucet. 

Dripping from a faucet is also well recognized as a classical model for a chaotic system 
\cite{Shaw1984}. Coullet et al.\cite{coullet2005hydrodynamical} proposed a hydrodynamics based mechanistic model to capture the chaotic behavior of droplet formation from a faucet. Further, this has also been well studied through experiments and numerical simulations 
\cite{kiyono1999dripping, ambravaneswaran2000theoretical}. 
The dripping dynamics of a faucet are controlled by the flow rate, surface tension of the fluid, viscosity and gravitational acceleration \cite {clanet1999transition, ambravaneswaran2000theoretical}. The different regimes of dripping are realized by modulating the Weber number (ratio of inertia and surface tension), $We = \rho U^2 R_n/\sigma$; where $\rho$ is the density of the liquid, $U$ is the velocity at the inlet, $R_n$ is the nozzle radius and $\sigma$ is the surface tension coefficient).  
Dynamics of droplet formation at a nozzle can be divided into three phases: (i) growth of a pendant droplet attached to the nozzle (ii) neck formation and pinch-off of the droplet (iii) retraction of the left over column of liquid at the nozzle. Further, dripping dynamics is usually characterized by the size of the droplet ($D_d$), length of the left over column of liquid ($L_d$) at the time of droplet pinch-off and the time period between successive droplet detachments. Different dripping regimes, depending upon the viscosity of the fluid, have been observed such as  periodic dripping with droplets of the same size (P1), periodic dripping with formation of intermittent satellite droplets (P1S),  dripping with droplets of two different sizes and breakup lengths forming alternatively (period-2 or P2 regime) and similarly period-4 regime and so on. Complexity of droplet formation and breakup, characterized by $D_d$ and $L_d$, continuously increases with increase in the flow rates and eventually chaotic dripping regime is observed where droplet size and length of the breakup vary with each droplet formation. Beyond a critical Weber number, $We_{d-j}$,  a sharp transition is identified from  dripping regime to `jetting' regime in Newtonian Fluids where with an increase in the flow-rate a sharp increase in $L_d$ is observed. In the jetting regime droplet formation occurs at the tip of a slender liquid column. The critical Weber number for dripping to jetting transition for different fluids has been well studied \cite{rubio2018dripping, rubio2013thinnest}.  Ambravaneswaran et al.\cite{ambravaneswaran2004dripping}  studied the effect of viscosity on the dripping to jetting transition using axisymmetric 1D equations derived in Ref. \cite{eggers2015singularities}. They showed that the dripping to jetting transition at sufficiently high viscosity occurs without the dripping regime going through the chaotic dripping regime and jetting occurs directly after the periodic regime, $P1$, in contrast to the dripping-jetting transitions proposed by the reduced order models of Coullet et al.\cite{coullet2005hydrodynamical} and Clanet et al.\cite{clanet1999transition}, where jetting is expected to occur after the chaotic dripping regime. Furthermore, based on the relative time scales corresponding to the convective flow, capillary viscous and inviscid times scales,  Ambravaneswaran et al.\cite{ambravaneswaran2004dripping} proposed the following scaling for $We_{d-j}$ using  Ohnesorge number $Oh = \sqrt{We}/Re$ (where $Re = \rho U R_n/\mu$) based on the liquid viscosity ($\mu$): (i) $Oh \ll 1$, $We_{d-j} \sim O(1)$ indicating that for low viscosity fluids, dripping to jetting transition has only weak dependence on the liquid viscosity, (ii) $Oh \sim O(1)$, $We_{d-j} \sim Oh^{-6}$ suggesting a very steep decrease in the critical Weber number with an increase in the viscosity and (iii) $Oh \gg 1$, $We_{d-j} \sim Oh^{-2}$ for highly viscous fluids. We note that these scalings which are in good agreement with the simulations \cite{ambravaneswaran2004dripping,subramani2006simplicity} suggest a decrease in the required flow rate (lower $We_{d-j}$) for the dripping-jetting transition with an increase in viscosity.

Rubio-Rubio et al. \cite{rubio2018dripping} further investigated the effect of viscosity on the critical Weber number for the dripping to jetting transition for different Bond numbers $Bo$ (ratio of gravitational force and surface tension force). Dripping dynamics (length of the liquid column at the droplet breakup, $L_d$, and the size of the droplets, $D_d$) remain independent of the Bond number for low flow rates. However, $We_{d-j}$ decreases with an increase in the Bond number. Moreover, $L_d$ in the jetting regime decreases significantly with an increase in $Bo$.
Zhang and Basaran \cite{zhang1996dynamics} studied the effect of an electric field, in addition to gravitational forces, on the dripping dynamics. They showed that with an increase in the applied electric voltage, elongated peer-shaped droplets form and dripping to jetting transition occurs at lower flow rates. Further, Wang et al.\cite{wang2021dynamics} showed that electric field induced oscillations of the meniscus can significantly alter the dripping dynamics.

It is interesting to note that there exits a hysteresis in these transitions where the critical Weber number identified is different as the flow regime transitions from a dripping to a jetting regime versus jetting to a dripping regime. Clanet et al.\cite{clanet1999transition} and Ambravaneswar et al. \cite{ambravaneswaran2004dripping} showed that the critical flow rate for jetting to dripping is smaller than that for dripping to jetting, that is $We_{j-d} < We_{d-j}$, where $We_{j-d}$ is the critical Weber number for jetting to dripping transition. Rubio-Rubio et al. \cite{rubio2013thinnest} studied the jetting to dripping dynamics for a range of viscosities. In order to study jetting to dripping transition, they performed a global linear stability analysis, accounting for weakly non-parallel effects and including the effects of viscosity and axial curvature. Based on the stability analysis, jetting to dripping transition can be determined by the border of the regions of absolute and convective instability. In particular,  the role of the gravitational stretching of the liquid jet on the transition from jetting to dripping was highlighted.

In the dripping regime, the retracting column of liquid, formed during droplet detachment, interacts with the incoming flow from the nozzle, and thus this process intricately dictates the formation of the successive droplets.  At low inlet flowrates, the retracting liquid column's dynamics is relatively faster and the initial pendant droplet profile at the nozzle stabilizes before any significant growth of the next droplet begins, thus resulting in a more consistent initial condition for each droplet and therefore periodic formation of droplet. With an increase in the inlet flowrate at the nozzle, droplet breakup occurs earlier and the incoming flow interacts with the retracting liquid column resulting in variations in the initial conditions for successive droplets resulting in period doubling events with period-2, period-4 and eventually chaotic dripping \cite{zhang1995experimental, zhang1999dynamics, ambravaneswaran2002drop}. Umemura \cite{umemura2016self} showed that growth of the short capillary waves travelling upstream on the retracting liquid column leads to the formation of neck on the liquid column resulting in satellite droplet formation events. Hoeffner \cite{hoepffner2013recoil} showed that the recoil dynamics of a liquid column may prevent an imminent pinch-off, thus delaying droplet formation and resulting in the formation of a liquid jet.

Transitions from dripping to jetting and jetting to dripping have been studied in the quasi-steady states, while the temporal dynamics aspects have been ignored in determining the critical Weber number for these transitions. In the present work, we perform experiments to study the droplet formation from a nozzle by progressively increasing the flowrate to obtain jetting regime and subsequently decreasing the flowrate to study transition from jetting to dripping regime. In particular, we focus on the transient evolution from a given initial condition (startup condition) to show that beyond a critical $We_{d-j}$, jetting is achieved only after several droplet formations have occurred in the dripping mode. With an increase in $We$ in the neighborhood of $We_{d-j}$, the number of droplets forming in the dripping regime before jetting is achieved decreases. In contrast, during the jetting to dripping transition, transient dynamics shows a sudden change. Using the reduced order 1D model of Eggers \cite{eggers1994drop, ambravaneswaran2000theoretical, shukla2020frequency}, we show that this dynamics is also captured in the simulations. Further, we show that the dripping to jetting transition and vis-a-versa can be altered significantly by introducing velocity perturbations at the nozzle. 

The paper is organized as following. Section \ref{sec:Experimental} presents the experimental setup and observations. Section \ref{sec:Discussion} focuses on demonstrating the underlying mechanisms using 1D axisymmetric and  using volume-of-fluid method in the two-phase flow open source code Gerris \cite{popinet2003gerris,popinet2009accurate,tomar2010multiscale}). In particular, the effect of viscosity (Kapitza number) and velocity perturbations at the inlet on the dripping-jetting and jetting-dripping transitions have been discussed. Finally, important conclusions from the study are presented in \ref{Sec:Conclusions}.

\begin{figure}[ht!]
  \centering
  \captionsetup{width=\linewidth,justification=justified}
  \includegraphics[width=1\textwidth]{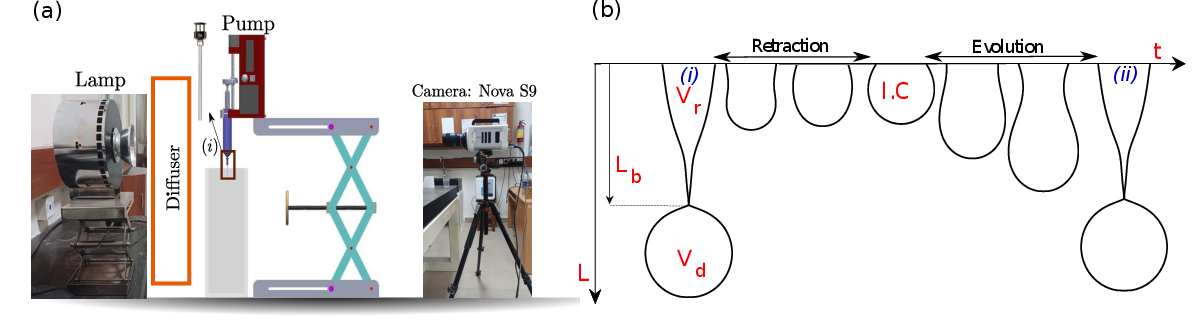}
  \caption{A schematic of the experimental setup with the various modules labeled in the diagram. 
  Droplets are generated at the nozzle ($i$) into a rectangular chamber and fluid is pumped  using a syringe pump mounted on a stand. Imaging is achieved using a high-speed camera (Photron Nova S9) and an LED light source is used for illumination. A diffuser sheet is used for providing a uniform backlighting from the LED for shadowgraphy. (b) Stages of droplet evolution between successive droplet breakups are shown, with parameters indicating the drop length ($L$), breakup length ($L_b$), and initial condition (I.C) between retraction and evolution, where the droplet volume ($V_d$) separates from the leftover liquid column, leaving behind the residual volume ($V_r$), respectively.  }
  
 
  \label{fig:testrig}
\end{figure}
 

\begin{table}[htbp]
\centering
\small
\begin{tabular}{lccccccc}
\hline
\textbf{Solvent} & $\rho$ (kg/m$^3$) & $\mu$ (Pa·s) & $\sigma$ (N/m) & $Ka$ & $Bo$ & $We$ Range & $U_m$ Range \\
\hline
PEGPG 45\% & 1051 & 0.130 & 0.0360 & 1.460 & 0.115 & 0.0028–4.24 & 0.05–2.06 \\
Glycerol 20\% & 1061 & 0.007 & 0.0695 & 0.048 & 0.060 & 0.0015–1.46 & 0.04–1.21 \\
Glycerol 60\% & 1153 & 0.026 & 0.0666 & 0.181 & 0.067 & 0.0025–2.87 & 0.05–1.69 \\
Silicone Oil & 960 & 0.048 & 0.0206 & 0.330 & 0.183 & 0.0049–5.57 & 0.07–2.36 \\
\hline
\end{tabular}
\caption{Properties and parameter ranges for the fluids used. Weber number and $U_m = \sqrt{We}$ are computed for flow rates varying from 1 to 30 ml/min through a 1.26 mm diameter nozzle.}
\label{Tab:Fluid}
\end{table}

\section{Experimental observations} \label{sec:Experimental} 

The experimental setup shown in figure \ref{fig:testrig} is used to study the transient dynamics between dripping to jetting regime of Newtonian liquid. The present setup includes a syringe pump (New Era Pump Systems, NE-1000) for precise control of the flow rate, a standard syringe coupled with a hypodermic needle of outer diameter $D_n=$1.26 mm and an assembly comprising a rectangular tank  (150 x 150 mm and with a depth of 500 mm) for collecting the dispensed fluid. The dimensions of the rectangular tank are chosen so as to prevent the ambient air flow to affect the dripping dynamics and is sufficiently wide to not influence the dripping dynamics physics. Photron Nova S9 high-speed camera is used for imaging to capture the drop breakup events and other transient dynamics accurately. To study the transient dynamics between dripping to jetting regime near the critical flow rates, the syringe pump is operated in two modes, namely, sweeping and step-wise. In the sweeping mode of operation, flow rate is changed (increased or decreased) at specified intervals of time. This allows us to study the hysteresis between dripping to jetting and jetting to dripping transitions. In the step-wise operation, experiments were conduced at a fixed flow rate to study evolution of successive pinchoff conditions. This mode of operation allows us to study both the dripping faucet dynamics and temporal evolution of successive droplets at specified flow rate. Such mode of operations have been discussed by Ambravaneswaran et al. \cite{ambravaneswaran2000deformation}.  The quasi-steady states conditions presented by Clanet (1999), Rubio-Rubio (2018) \cite{clanet1999transition, rubio2018dripping}, illustrate that the critical flow rate ($Q_c$), at which dripping-to-jetting transition occurs, is a function of fluid properties, gravitational acceleration and the nozzle diameter. In this study, a step-wise mode of operation is used to study the transient evolution from a given initial condition (startup condition) to show that beyond a critical flow rate, jetting is achieved only after several droplet pinchoff events in the dripping regime have occurred before jetting is observed. At each flow rate, experiments were conducted at the critical flow rates to observe the temporal evolution of the droplet. These experiments were conducted for different fluid properties by changing the $\%$ Glycerol-water solutions and also with high viscous silicone oil and a 45$\%$ aqueous solution of PEGPG.\\ 
 \begin{figure}
  \captionsetup{width=\linewidth}
  \captionsetup{justification=justified} 
    \includegraphics[width=0.7\textwidth]{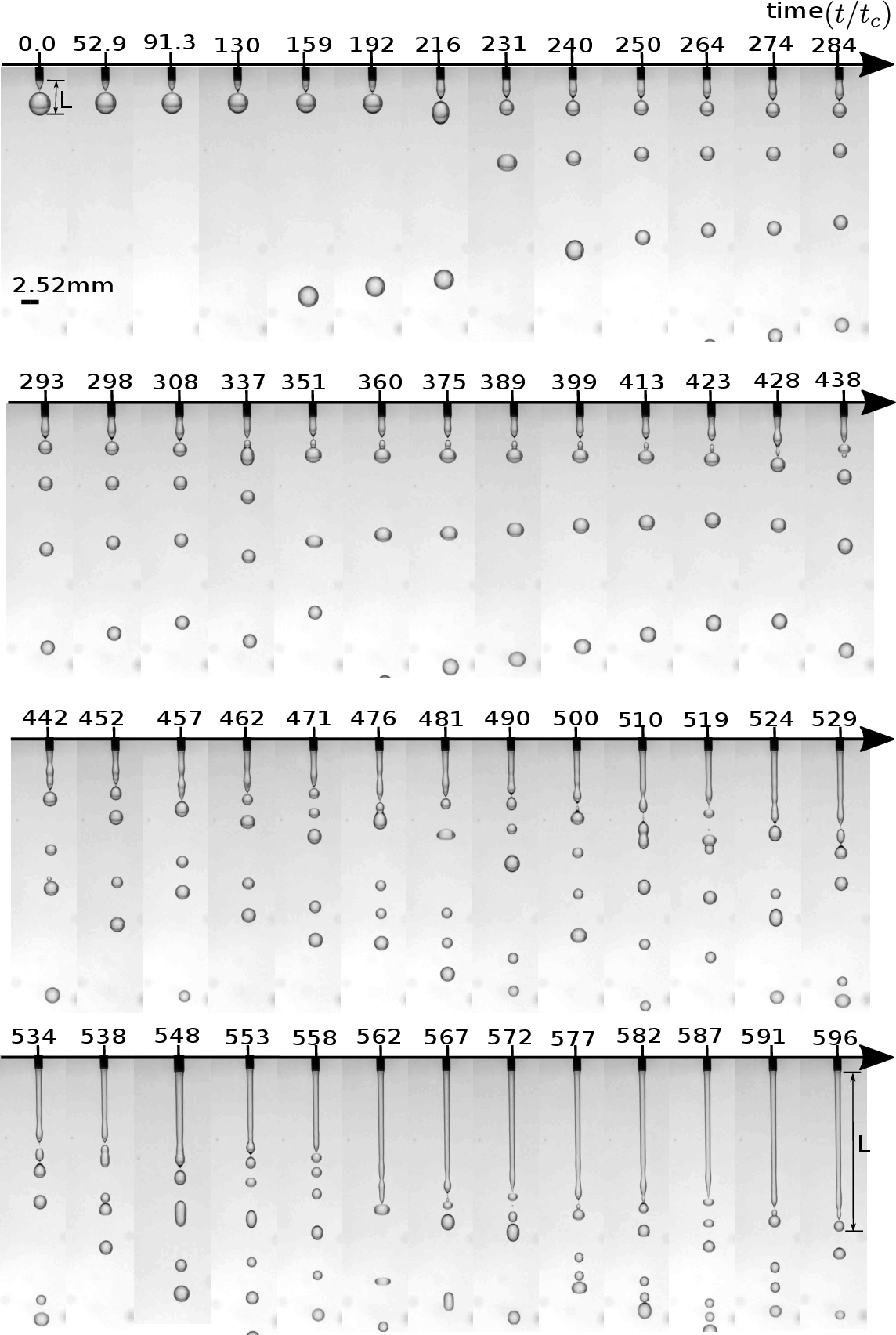}%
 \caption{\label{fig:breakup_profiles} Drop formation dynamics for $U_m=0.88$,  $Bo=0.0674$ and $Ka=0.181$ is shown. Time stamp $t/t_c$ is 
 marked in the figure with $t/t_c = 0$ corresponding to when the syringe pump is switched on for the given flow-rate.  The snapshots shown correspond to the 
 pinchoff events and the corresponding the pinchoff length, $L_b$, is noted.}
\end{figure}
The transient evolution of the droplets is visualized using the shadowgraphy module employing a high-speed camera (Photron Fastcam Nova SA9). The camera is interfaced with a computer with an i7 processor running the image acquisition software. The setup is illuminated by an LED source with a diffuser plate. As the current study is focused explicitly on the dripping faucet and the transition from dripping-to-jetting and jetting-to-dripping, we have chosen a larger vertical resolution using 256 pixels in width and 1024 pixels in height. A typical experimental run here includes recording the evolution of the  droplets at a constant flow rate and at a rate of 6400 frames per second until quasi-periodic conditions of dripping are obtained. The high-resolution images of the liquid jet were post-processed to track the droplet evolution using droplet tip length ($L$), successive droplet pinchoffs using the pinchoff length ($L_b$) and droplet diameter ($D_d$) (see figure \ref{fig:testrig}b). Figure\ref{fig:testrig}b shows one droplet formation cycle with retraction of the left over liquid column after breakup, I.C being the initial condition for the growth of the next droplet and corresponding evolution stage for droplet formation. Canny edge detection and dilation (of the image to enhance the contrast of the edge) algorithms were used to post-process the images. Using the dilated image so obtained, key parameters of the liquid jet such as the edges of the jet and droplet contours were detected. The minimum neck radius is calculated by finding the row with the smallest pixel count in the dilated edge-detected image and is then used to locate the tip of the jet ($L$). This algorithm is applied to successive images captured during the experiments. If there is a sudden jump in the jet tip length, this is identified as the pinchoff length of the jet ($L_b$). Similarly, the droplet diameter is calculated by identifying the contour of the droplets. These contours are identified from the frame in which the pinchoff length is detected. The contour of the droplets is tracked for the next 50 frames to average out the droplet diameter oscillations. The diameter of each droplet is calculated from the contour area ($A$) using the formula $D = \sqrt{4A/\pi}$.\\ 
The droplet evolution from a faucet is essentially governed by the following parameters: the transport properties of the liquid, the size of the nozzle $D_n$, the flow rate $Q$, and the gravitational acceleration $g$. The corresponding three non-dimensional parameters that govern the liquid jet evolution are Kapitza number ($Ka$), Weber number ($We$) and Bond number ($Bo$). The Kaptiza number is given by \(
Ka =3 \nu \left( \rho^3 g/\sigma^3 \right )^{1/4} = 3(l_{\nu}/l_c)^{3/2} \), and represents the ratio of the viscous forces to capillary length scales. Here, the capillary length scale is $l_c= \sqrt{\sigma/ \rho g}$ and the viscous length scale, $l_{\nu} = \nu^{2/3} g^{-1/3}$, emerges from the interplay between the gravitational and viscous forces. The Weber number is defined as $We = \rho U^2 R_n/\sigma$ is the ratio of inertial to capillary forces. The non-dimensional inflow velocity can be defined as the ratio of inlet velocity $U$ to the capillary velocity:  $U_m = U/U_c =\sqrt{We}$ where $U_c=\sqrt{\sigma/\rho R_n}$. The Bond number, $Bo=\rho g R_n^2/\sigma$, relates the gravitational forces to the surface tension forces. Additionally, the Ohnesorge number $Oh=\sqrt{We}/Re$, is the ratio of the viscous-capillary time ($t_{\mu}=\mu R_n/\sigma$) to the inertial-capillary time  ($t_c=\sqrt{\rho R_n^3/ \sigma}$) \cite{villermaux2013viscous}. Here, we define the Ohnesorge number as $Oh=Ka/3Bo^{1/4}$ (\cite{ambravaneswaran2004dripping,clanet1999transition, anthony2023sharp, rubio2018dripping}). In the  experiments presented, we use a variety of Newtonian liquids, the properties of which are given in table \ref{Tab:Fluid}. \\
Figure \ref{fig:breakup_profiles} shows interface profiles for the inlet velocity $U_m =0.88$ (corresponding to the dimensional inlet velocity $U = 0.30$ m/s) for Glycerol-water mixture with $Ka=0.181$ (see Table \ref{Tab:Fluid}) and $Bo = 0.0674$. Interface profiles are shown along with a time stamp ($t/t_c$) at the top of each profile. The timeline mentioned in the figure starts from the first droplet pinchoff event (labeled $t/t_c=0$). Interface profiles are further shown for successive droplet pinchoff events at $t/t_c = 52.9$, representing the second pinchoff. With each droplet formation, the droplet size decreases and the pinchoff length increases (defined as the length of the liquid column at the pinchoff point as marked in the figure). The pinch-offs occur at irregular time intervals, $t/t_c = $ 52.3, 91.3, 130,159,..., thus indicating chaotic dripping with irregular variation in pinchoff length and droplet size. We note that the droplet formation time decreases, although not in a monotonic way. For instant, around $t/t_c = 52.9$, the time interval between droplet formations was $\sim 38.4$, at $t/t_c = 293$, time interval reduces to $\sim 5-10$units and at around $t/t_c = 534$ it became more consistent at about 5 units. The pinchoff length along with the droplet size also varies with each pinchoff. As time progresses, the variation in length at pinchoff increases. For example compare the interface profiles at $t/t_c = 442$ with that for $t/t_c = 284$ and $t/t_c = 52.9$. We notice an overall increase in the pinchoff column length and a decrease in the droplet size with each subsequent droplet formation. 
 
This evolution process indicates that the transition 
from dripping to jetting generally occurs smoothly over time after many successive droplet formations and not abruptly. The interface profiles suggest formation of capillary waves that travel upstream and set stage for the next pinchoff event.
Since the ratio of inflow velocity to capillary velocity is much smaller ($U_m=U/U_c \ll 1$), gravity dominates the droplet evolution in the dripping regime. However, for $U_m \gg 1$, the Rayleigh-Plateau regime is expected \cite{safronov2021fast}. Here, we show that transient evolution is observed for an inflow velocity comparable to the capillary velocity  ($U_m = 0.88 < 1$).\\ 
 \begin{figure}[ht!] 
  \centering
   \captionsetup{width=\linewidth}
  \captionsetup{justification=justified} 
    \includegraphics[width=0.85\textwidth]{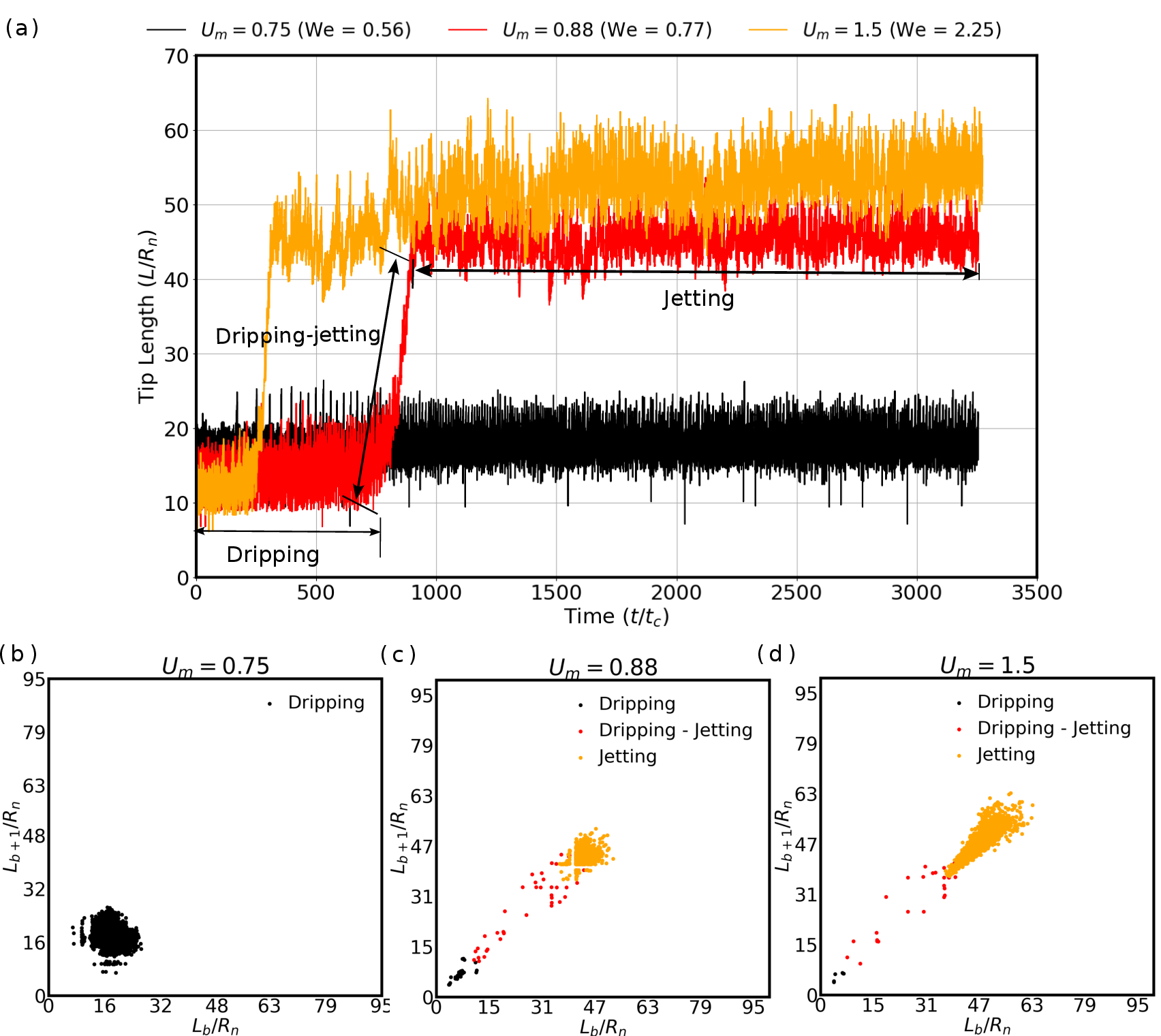}
\caption{\label{fig:Length_time_flowrate_variation} (a) Temporal evolution of length of jet at different inflow velocities. The plots demonstrate an increase in length as time progresses for a given flow rate in Dripping to jetting transition regime $U_{m_dj}$ for $U_m\geq 0.88$. As the Weber number increased to $1.5$, the transition to the jetting has accelerated, occurring at  $t/t_c=312$ compared to $t/t_c=888$ at $U_m=0.88$. Poincare maps of breakup lengths $L_b$ for increasing flow rates, showing (b) dripping regime, (c) Dripping to jetting transition, (c) accelerated jetting regime by $Ka=0.181$, $Bo=0.0674$. }
    \end{figure}  
 Before we proceed to discuss transition mechanisms between dripping and jetting regimes, it is important to note the characteristics of the droplet formation process that delineates the dripping regime from the jetting regime. 
 A significant change in the pinchoff or break-off length ($L_b/R_n$) and the droplet size distribution ($D_p/D_n$) are used to describe the differences between these regimes. Figure \ref{fig:Length_time_flowrate_variation}a shows variation in the length of the liquid column (tip length) attached to the nozzle, $L/R_n$, with time.
 For $U_m = 0.75$ (black contour), although the length of the liquid column attached to the nozzle varies chaotically, it does not undergo a significant change. However, for $U_m = 0.88$, after several droplet formations at shorter pinchoff lengths, $L/R_n$ rapidly increases to a significantly higher length ( $\sim 5$ times). We can thus claim that for $U_m = 0.88$ jetting regime is achieved in the quasi-steady state. In contrast, for $U_m = 1.5$, the number of droplet pinchoffs before jetting regime is achieved are very few ($\sim 10$ droplet formations in the dripping regime). The corresponding 
  Poincare plots for pinchoff lengths $L_b/R_n$ at three different speeds $U_m = 0.75, 0.88$, and $1.5$ are shown in Figures \ref{fig:Length_time_flowrate_variation}b,c, and d, respectively. Here, Poincare plots are used to study periodicity in the pinchoff length $L_b/R_n$. The three different regimes that we use for classifying the droplet formation are: (i) dripping regime shown with black dots, (ii) dripping-jetting regime where length increases from dripping towards jetting regime (red dots) and (iii) jetting regime where droplet formation occurs from a long liquid column called as jet (orange symbols). We have marked these regions in  \ref{fig:Length_time_flowrate_variation}a for $U_m = 0.88 $. The Poincare plots illustrate a progressive change with an increase in the inlet velocity. For $U_m = 0.75$, all the pinchoff lengths are concentrated at about $L_b/R_n = 16$ (similar to experimental observation in \cite{sartorelli1994crisis, kiyono1999dripping}). With an increase in $U_m$ to 0.88, there are some dots in the dripping-jetting transition regime with slightly shorter pinchoff lengths ($\sim 15$), and $L_b/R_n \sim 47$ in the quasi-steady jetting state (marked in orange). The transition occurs from dripping to jetting through a continuous increase in the pinchoff length marked with red dots. For $U_m = 1.5$, there are fewer black dots ($\sim 10$ droplet formations in the dripping regime) indicating only a few droplets in the dripping regime and a much sharper transition to jetting with very few Poincare points in the intermediary pinchoff lengths. Further, in the jetting regime for $U_m = 1.5$, we note a much wider variation in the pinchoff lengths. We also note here that the length of the jet increases with $U_m$. This is in agreement with the observations of Lin and Ritz\cite{lin1998drop} where initially an increase in the inlet Reynolds number leads to an increase in the breakup or jet length and subsequently beyond a critical Reynolds number (inlet velocity) the jet length decreases during the atomization regime. 
  The transition between dripping and jetting regimes is usually marked by the inlet $We$. For $U_m = 0.75$, $We = 0.56$ and for $U_m = 0.88$ $We = 0.77$. 
The scaling suggested by Ambravaneswaran \cite{ambravaneswaran2004dripping} based on the axi-symmetric simulations for the critical $We$ for transition yields $We_{d-j} \sim Oh^{-2} = 0.14$. As discussed earlier, transition from dripping to jetting regime is usually characterized by a sudden change in the pinchoff length and the critical $We_{d-j}$ is marked by the inlet velocity at which jetting regime ensues from the beginning (\cite{clanet1999transition, rubio2013thinnest, trettel2020reevaluating}) unlike the transition discussed here which occurs after several droplet formations in the dripping regime and jetting regime is achieved later in the quasi-steady state.

  \begin{figure}[ht!] 
  \centering
   \captionsetup{width=\linewidth}
  \captionsetup{justification=justified}  
  \includegraphics[width=0.95\textwidth]{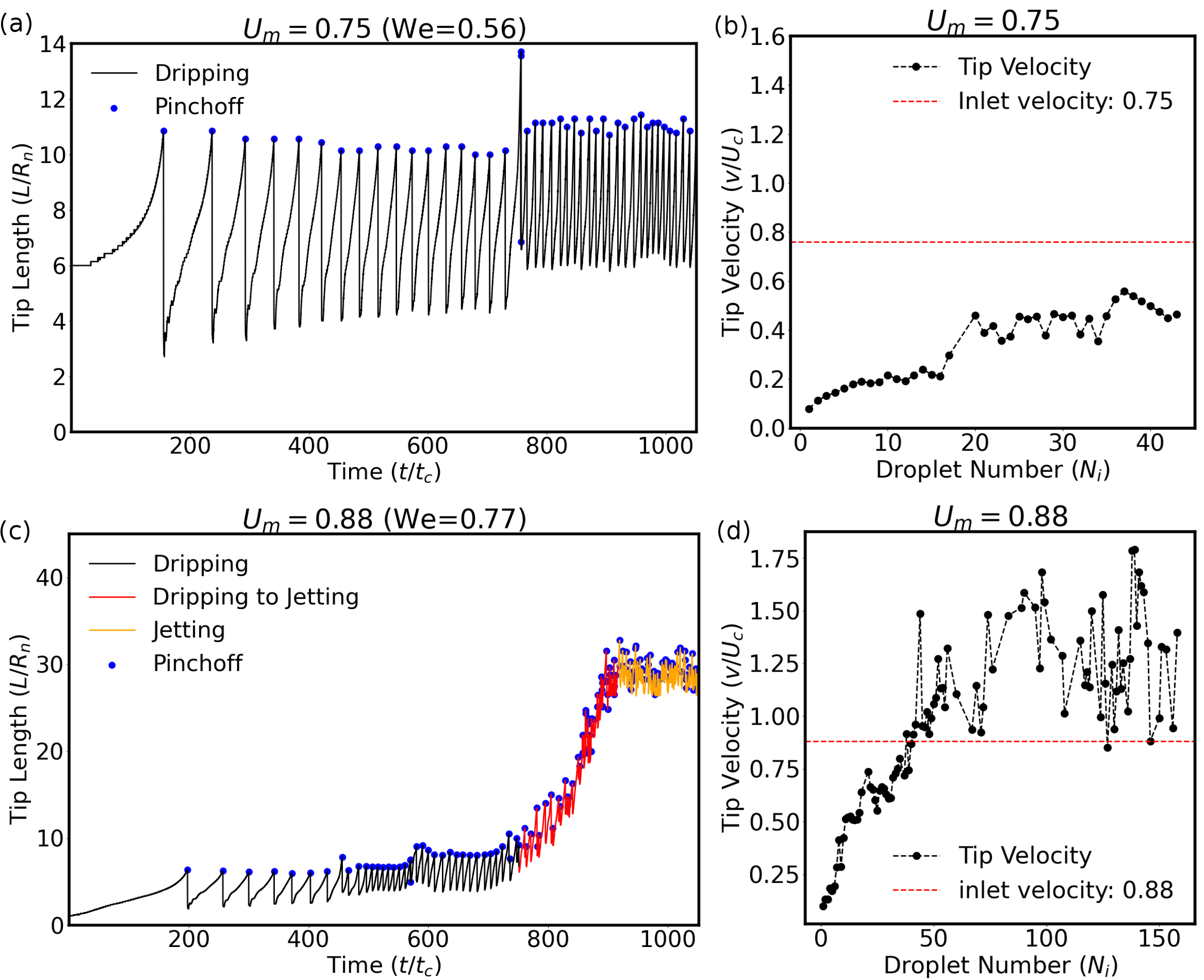}
  \caption{\label{fig:Tip_velocity} (a,c) Variation in droplet tip location and pinchoff location from the nozzle as a function of time is showing dripping regime and dripping to jetting transition regime through the increase in velocity from $U_{m_d}$ to $U_{m_j}$. (a) The periodicity of pinchoff length shifts with the transition to a new state of periodicity. (b) Highlights the increase in the tip velocity between successive droplets after 720, where droplet length increases yet remains below the inlet velocity condition ($v/U_c = 0.76$). (c) Captures the transition to jetting at 780, 888 through increase in length, with saturated length in the jetting regime shown in the orange line. (d) Indicates that for upto 50 droplets, the jet velocity remained below the inlet velocity of 0.88, transitioning to jetting as the tip velocity oscillates above the inlet velocity for the primary pinchoff.}
  \end{figure}  
 
Figure \ref{fig:Tip_velocity} shows variation in tip length $L/R_n$ (zoomed in view of the figure \ref{fig:Length_time_flowrate_variation}a) and tip velocity ($v/U_c$) between successive pinchoff conditions for two different inlet velocities. Tip length variation can be categorized into two stages. In the first stage after the pinch-off of a droplet (marked as blue dots in the figure), tip retracts and subsequently after the retraction phase as the droplet grows the tip of the liquid column attached to the nozzle increases and undergoes a rapid increase just before the pinchoff. The pinchoff for $U_m = 0.75$ shown in figure \ref{fig:Tip_velocity}a consistently occurs around $L/R_n \sim 10.5$ albeit not periodically and at $t/t_c \sim 800$ undergoes a change resulting in the droplet formation occurring at a higher frequency (23 Hz) compared to   (17 Hz initially) with a slightly longer pinchoff length $\sim 11.3$. Figure \ref{fig:Tip_velocity}b shows the tip velocity at the time of pinchoff. For the first droplet the tip velocity is rather slow $\sim 0.2 U_c$ where  $U_c = \sqrt{\sigma/\rho R_n}$) is the capillary velocity. Tip velocity increases with every subsequent droplet formation. However, the tip velocity never exceeds the velocity at the inlet indicating that the left over liquid column after droplet pinchoff gets sufficient time to recoil. The tip velocity for the droplet pinch-off at which sudden transition occurs at around the inlet velocity. We note that the jump in tip length also corresponds to a sudden increase in tip velocity. For $U_m = 0.88$, figure \ref{fig:Tip_velocity}c shows a rapid variation in the tip length and after several droplet pinchoffs in the dripping regime (marked with solid black line), droplet formation transitions into jetting regime (marked with solid orange line) through a transition phase marked in red. The frequency of droplet formation (162 Hz) is significantly higher compared to that for $U_m = 0.75$. The corresponding tip velocity variation shows an increase with every droplet formation and after about 50 droplet pinchoffs exceeds the inlet velocity of 0.88 (similar to the observation at higher inlet velocity conditions in \cite{rezayat2021high}). This also marks the beginning of the jetting regime. Thus, over a small change in the $U_m$ we note a significant change in the transient behavior of dripping dynamics.

\begin{figure}[ht!]
 \centering
 \captionsetup{width=\linewidth}
 \captionsetup{justification=justified} 
 \includegraphics[width=0.95\textwidth]{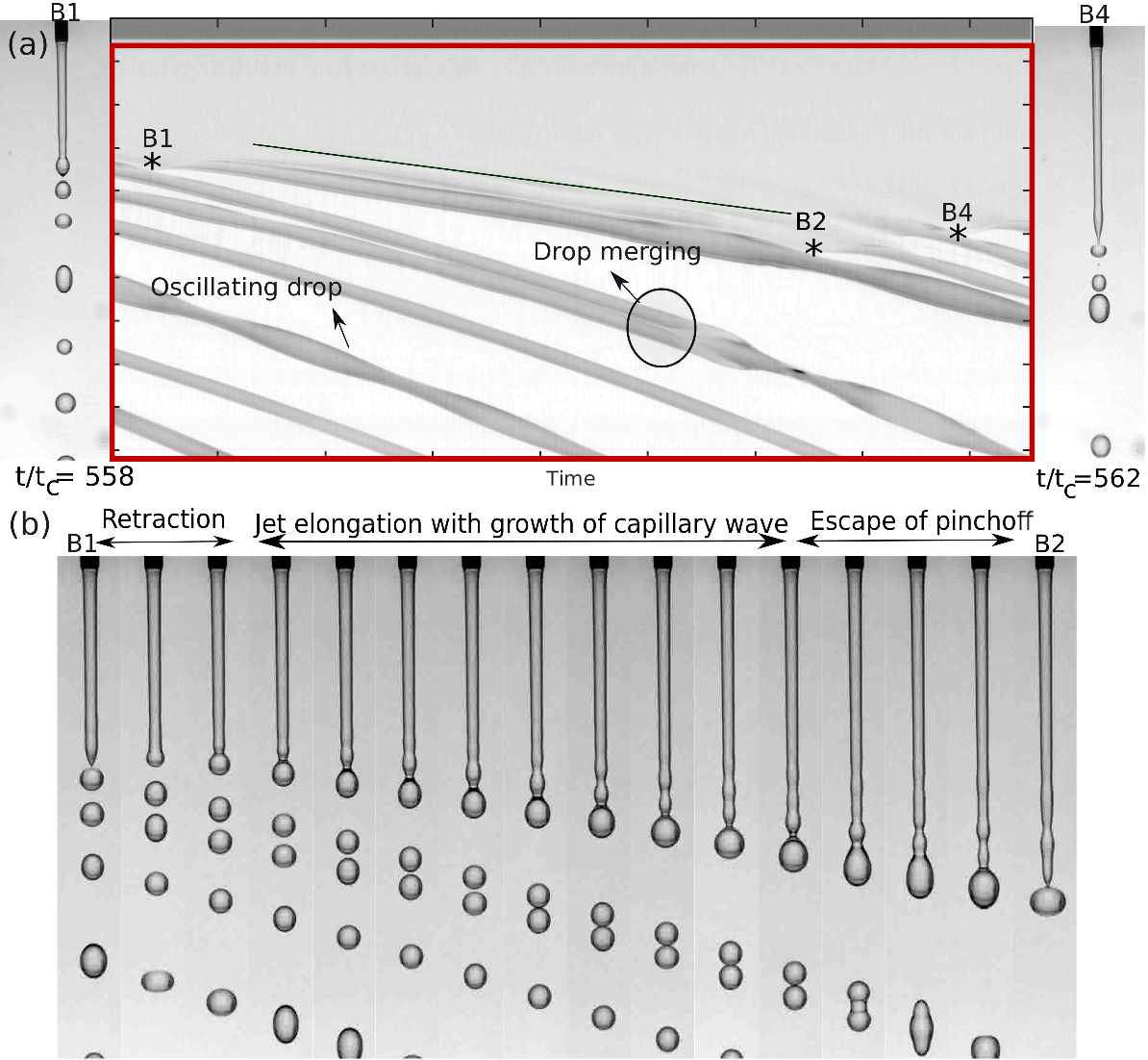}
 \caption{\label{fig:Escape_pinchoff} (a) Space-time evolution of between four pinchoff locations of the figure \ref{fig:breakup_profiles}. (b) liquid jet evolution indicates different events that occur as the jet length increases.}
\end{figure}

In order to understand the variation in the tip length from dripping regime towards jetting regime after several droplet formations, we examine the mechanisms behind the increase in the tip length.  Figure \ref{fig:Escape_pinchoff}a shows the spatio-temporal plot of the liquid jet for $U_m = 0.88$. Intensity in the figure indicates the thickness of the liquid in the radial plane at a given time and the variation in the thickness in the vertical direction can be used to obtain contiguous liquid regimes thus demarcating droplets from the liquid column attached to the nozzle.  In figure \ref{fig:Escape_pinchoff}a four successive droplet breakups have been shown marked as  B1, B2, B3 and B4. The interface profiles corresponding to B1 and B4 are shown adjacent to the spatio-temporal plot. During two consecutive droplet formations B1 and B2, the length of the liquid column (tip length) initially decreases indicated as the retraction phase in figure \ref{fig:Escape_pinchoff}b and subsequently increases as the droplet at the tip grows towards  pinchoff at B2. After droplet breakup at B1, during the retraction phase, capillary waves travel upstream of the liquid column towards the nozzle. These capillary waves generate a growing perturbation that results in the neck formation setting stage for the next breakup through the long wave breakup mechanisms similar to those described by \cite{umemura2011self} for inviscid liquids. We note here that during the growth of the droplet, oscillations are observed in the shape of the droplet from prolate to oblate. We believe that these oscillations result in pressure variations that eventually result in coarsening of the neck region as seen in figure \ref{fig:Escape_pinchoff}b and thus lead to a delay in the droplet breakup. We call this phase as the  escape pinchoff phase similar to that discussed in \cite{hoepffner2013recoil} for droplet formations from retracting ligaments. We note that these oscillations may also result in droplet merging events as marked in the spatio-temporal plot \ref{fig:Escape_pinchoff}a thus modifying the eventual droplet size distribution. 



\begin{figure}[ht!]
 \centering
  \captionsetup{width=\linewidth}
  \captionsetup{justification=justified} 
  \includegraphics[width=1\textwidth]{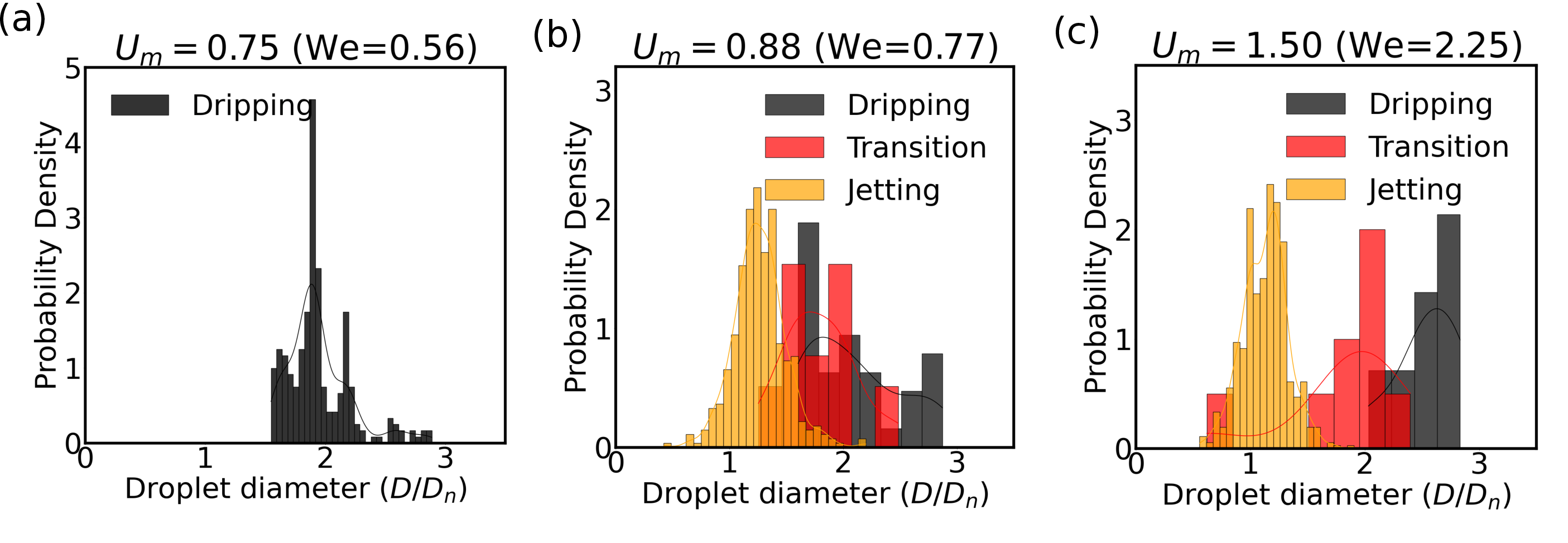}
  \caption{\label{fig:Droplet_size_distribution} The probability Density of droplet diameter distribution during dripping to jetting regime as the flow rate increases. (a) A right-sided distribution with droplet diameters beyond 2 is observed in the Dripping regime. (b) At $0.88$, during the dripping to jetting transition, droplet distribution centres symmetrically around a diameter of 1.5mm, whereas in the dripping regime, droplet sizes are close to 2. (c) At $1.50$, the distribution shifts towards smaller droplet sizes as the liquid values as the liquid pinchoff accelerates to jetting within 312.}
\end{figure}

 Figure \ref{fig:Droplet_size_distribution} shows the variation in the droplet size distributions in dripping, dripping-jetting and jetting regimes corresponding to the ones marked in figure \ref{fig:Length_time_flowrate_variation}. Different colors have been used to represent the different regimes at a given $U_m$. For $U_m = 0.75$, figure \ref{fig:Droplet_size_distribution}a shows that in the dripping regime the mean droplet size distribution is around two times the nozzle diameters (see \cite{li2016capillary}). The curve shown in the figure shows the kernel density estimate ($KDE$) of the distribution (see Appendix \ref{appB} for details).  The distribution is scattered between $1.6$ $\&$ $2.5$ with higher probability at 1.7. These variations in the droplet sizes in the chaotic dripping regime for $U_m = 0.75$ correspond to the variation in the tip length shown in figure \ref{fig:Tip_velocity}a. For instance an escape pinchoff event results in a droplet size of about 3.
 Figures \ref{fig:Droplet_size_distribution}b and c show the shift in the droplet sizes to smaller sizes below $D/D_n=2$ for $U_m=0.88$ and 1.5 in the jetting regime. After about 15 droplet pinchoffs the size of the droplets for $U_m = 0.88$ decreases (transition region marked in red) and shifts further to smaller droplet sizes in the jetting regime (marked in orange in the figure). Droplet sizes in the dripping regime for $U_m = 0.88$ are comparable to that for $U_m = 0.75$. For $U_m = 1.5$, droplet sizes in the jetting regime are much smaller compared to that for $U_m = 0.88$. These variations in droplet sizes directly correspond to the jet length, where longer lengths imply thinner jet radius at the tip and thus smaller droplets.
 
 \begin{figure}[ht!]
\centering
 \includegraphics[width=1\linewidth]{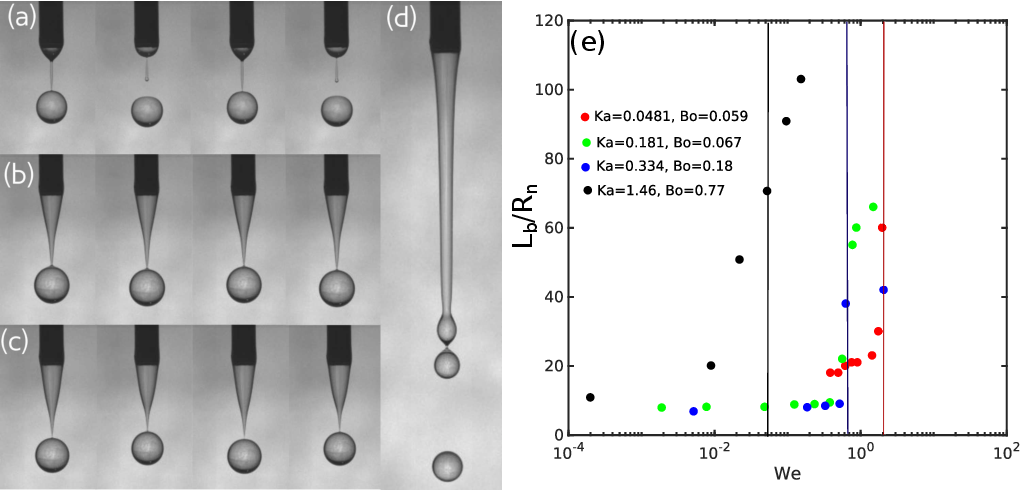}
\caption{\label{fig:regime} Effect of velocity on the droplet pinchoff length in he dripping faucet and jetting regimes are shown for  $Ka=0.33$ and $Bo=0.18$ (a) P1S1 at $We=0.0011$, (b) P1 at $We=0.07$, (c) P2 at $We=0.11$ and (c) jetting at $We=0.14$. (e) The effect Kapitza number $Ka$ on the Dripping to jetting transition as the Weber number is increased for different Bond numbers.}
\end{figure}

In figure \ref{fig:regime} dripping faucet and variation of mean breakup length are shown as a function of  $U_m$ for different Kapitza number, $Ka$, that is for Glycerol mixtures, PEGPG and Silicone oil, with different viscosities and surface tension coefficient given in table \ref{Tab:Fluid}. Figures \ref{fig:regime}a-d show the influence of Weber number on the dripping dynamics. At lower $We$, different periodic regimes such as P1S1, P1, P2 and jetting are observed as shown in figures \ref{fig:regime}a-d for $Ka=0.33$. With increase in $We$ from 0.0011 to 0.11, dripping regime is observed without any progressive increase in the liquid column length. However, at $We = 0.14$, the transient behavior in dripping-jetting transition is observed. Figure \ref{fig:regime}e shows variation in the breakup length $L_b/R_n$ with $We$ for different values of $Ka$ and $Bo$. Since in experiments we cannot vary $Ka$ and $Bo$, we would be performing systematic numerical experiment in the next section to delineate the effects of surface tension and viscosity. Here, in figure \ref{fig:regime}e we note that for lower $We$, where dripping is expected we obtain lower $L_b/R_n$ and with increase in $We$, the breakup length increases and eventually shows a distinct jump indicating transition from dripping to jetting regime. For different fluids i.e. different values of $Ka$ and $Bo$, the transition Weber number, $We_{d-j}$, is different. The bar at each of these steady state jetting regimes (shown only for cases where more than one droplet formation occurs before jetting is achieved) indicates the number of droplets forming before jetting is achieved corresponding to the data points in figure \ref{fig:regime}e.
We note that once the jet has formed, a further increase in $We$ increases the jet length further. For $Ka = 0.0481$, dripping to jetting transition occurs at a higher $We$ indicated by the red line and the critical $We$ for these transition decreases with increase $Ka$ (where $Bo$ also simultaneously increases). We note that transition for $Ka = 0.0481, 0.181, 0.334$ and $1.46$ occurs at $We_{d-j} = 1.98, 0.77, 0.63$  and $0.0528$, respectively. $We_j$ marks the jetting Weber number at which jet is formed without forming intermediate droplet formation cycles. In the thin band of ($We_{d-j}, We_j$), the transient dynamics involving formation of droplets is observed. We note that this transient dynamics is more prominent for higher $Ka$.

In what follows, we perform numerical simulations to further explore the parameter space to understand the underlying dynamics that result in such transient behavior during transition from dripping to jetting regime.

 \begin{figure}
 \centering
 \includegraphics[width=0.5\textwidth]{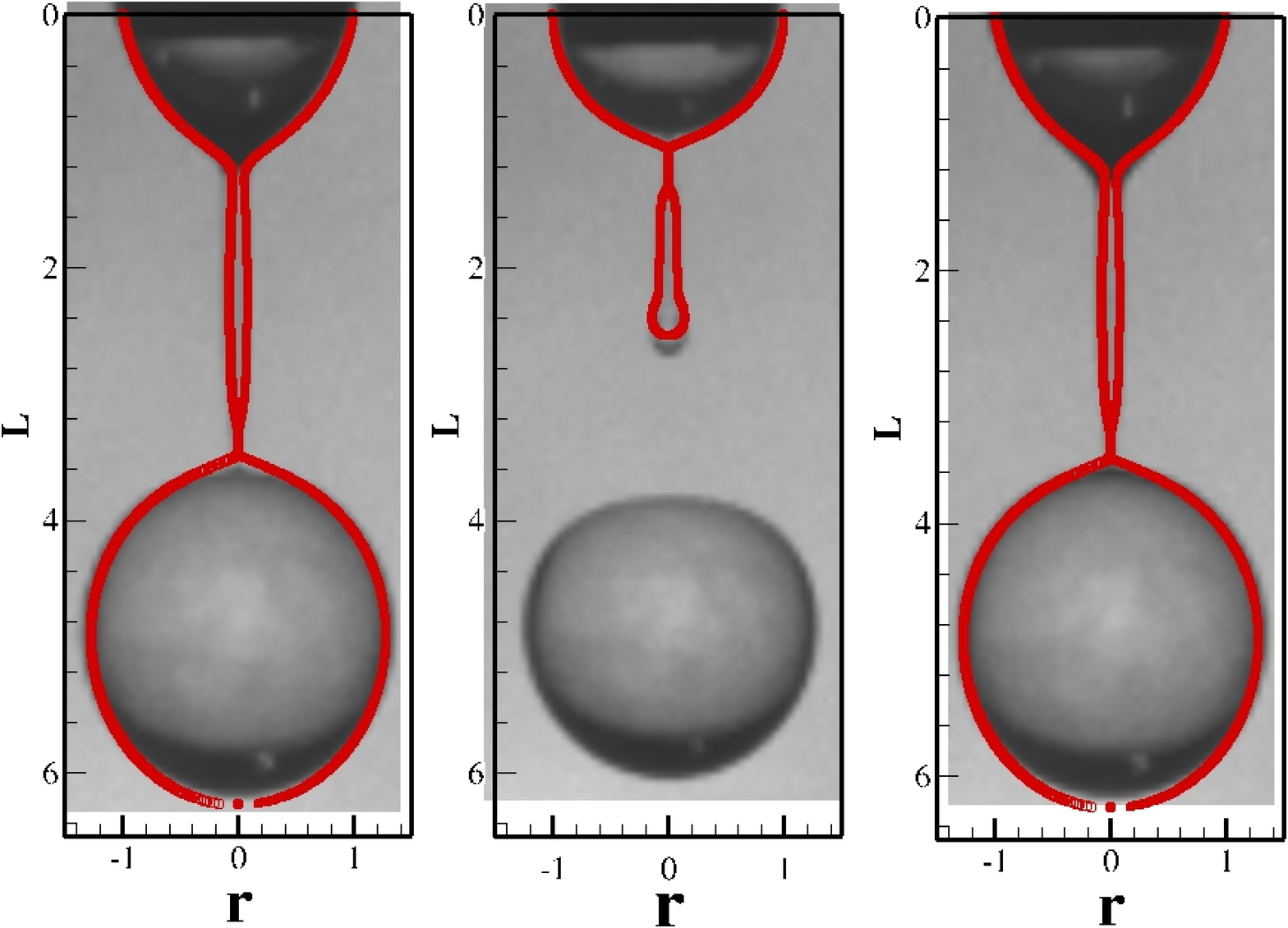}\\
 \captionsetup{width=\linewidth}\captionsetup{justification=justified} 
\caption{\label{fig:validation} Comparison of numerical simulations (red coloured interface)  with the experimental results for $Ka=0.181$, $We=0.0019$, $Bo =0.067$}.
\end{figure}

\section{Numerical Simulations}
\label{sec:Discussion}



In order to investigate the role of fluid properties in more detail, we perform numerical simulations using the slender jet equations proposed in \cite{eggers1994drop}.  Here we follow the method used in \cite{subramani2006simplicity, shukla2020frequency} by solving Eqns.\ref{eq:ND_KE} and \ref{ND_momentum} using an explicit finite difference method. The slender jet model accurately captures the dynamics of an axisymmetric liquid jet subjected to gravitational acceleration. The equations assume that the  surrounding gaseous medium does not influence the process of droplet formation at the nozzle. This assumption is justified given the large density and viscosity ratios between the gas and the liquid. Our full Navier-Stokes solutions are in good agreement with the reduced order model employed here and Subramani et al.~\cite{subramani2006simplicity} (see Appendix \ref{appA} figure \ref{fig:Gerris_Domain}c,d). During the evolution of the liquid jet and droplet pinchoff, we use the transformation $a = h^2$, where $h$ is the radius of the jet, to numerically simulate the pinch-off conditions similar to the one discussed in Isha et al. \cite{shukla2020frequency}. Here, $h$ is the jet radius, $v_0$ is the mean axial velocity, $Oh$ is the Ohnesorge number and $Bo$ is the bond number.

 \begin{equation}\label{eq:ND_KE}
    \dfrac{\partial a}{\partial t} +\dfrac{\partial  av_{0}}{\partial  z }= 0	 
\end{equation}
the non-dimensional form of the equation
\begin{equation}
\dfrac{\partial v_{0}}{\partial t} =-v_{0}\dfrac{\partial v_{0}}{\partial  z}  + {\dfrac{3Oh}{a}} \Biggr(\dfrac{\partial}{\partial z} a\dfrac{\partial v_{0}}{\partial z}\Biggr) - \dfrac{\partial}{\partial z} \left (  H\right)+ Bo 
\label{ND_momentum}
\end{equation}
Here, $H$ in the curvature term of the equation is defined as,
\begin{equation}
 H = \Biggr(\dfrac{\Big(2-\dfrac{\partial^2 a}{\partial  z^2}\Big)a+\Big(\dfrac{\partial a}{\partial z}\Big)^2}{2\Big(\dfrac{1}{4}\Big(\dfrac{\partial a}{\partial z}\Big)^2+a\Big)^{3/2}}\Biggr).
 \end{equation} 
 
We assess the efficacy of our numerical model through rigorous validation against both present experimental results as shown in figure \ref{fig:validation} for $Ka = 0.181$, $Bo = 0.067$ and $We=0.0019$. A hemispherical droplet is used as the initial condition for these simulations. Indeed, this simplification that does not capture contact angle variations of wetting fluids \cite{sedighi2021capillary}. But for simulating the non-wetting fluids, this condition allows examining the influence of fluid properties on the critical transition from dripping to jetting at a constant Bond number. The figure shows formation of three consecutive droplets. We note that the current simulations capture the shape and dynamics of the droplet formation, pinch-off of the droplet and also the formation of the satellite droplet, accurately. 
 

\begin{figure}[ht!]
   \centering
  \includegraphics[width=1\linewidth]{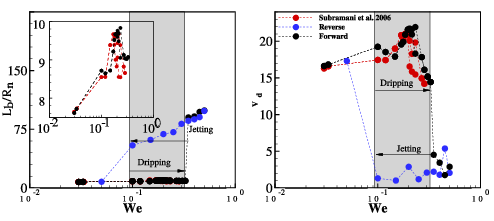}  
  \caption{\label{bifurcation} Illustration of dripping faucet and hysteresis effects near the dripping to jetting transition for a slender jet at $Ka=1.6$ and $Bo=0.33$. Panel (a) captures the transition sequence from the dripping faucet through periodic regimes P1 and P2, culminating in the jetting regime at $U_m=0.6$, characterized by a significant reduction in droplet volume to $V_b=5$. Conversely, under reverse flow rate conditions, a transition from jetting back to dripping is observed at $U_m=0.34$, where the pinchoff length decreases to $L_b/R_n=10$, and the corresponding droplet volume increases to $V_b=15$, highlighting the hysteresis behaviour in the transition between these regimes.}
 \end{figure}

\begin{figure}[ht!]
\centering
 \includegraphics[scale=0.5]{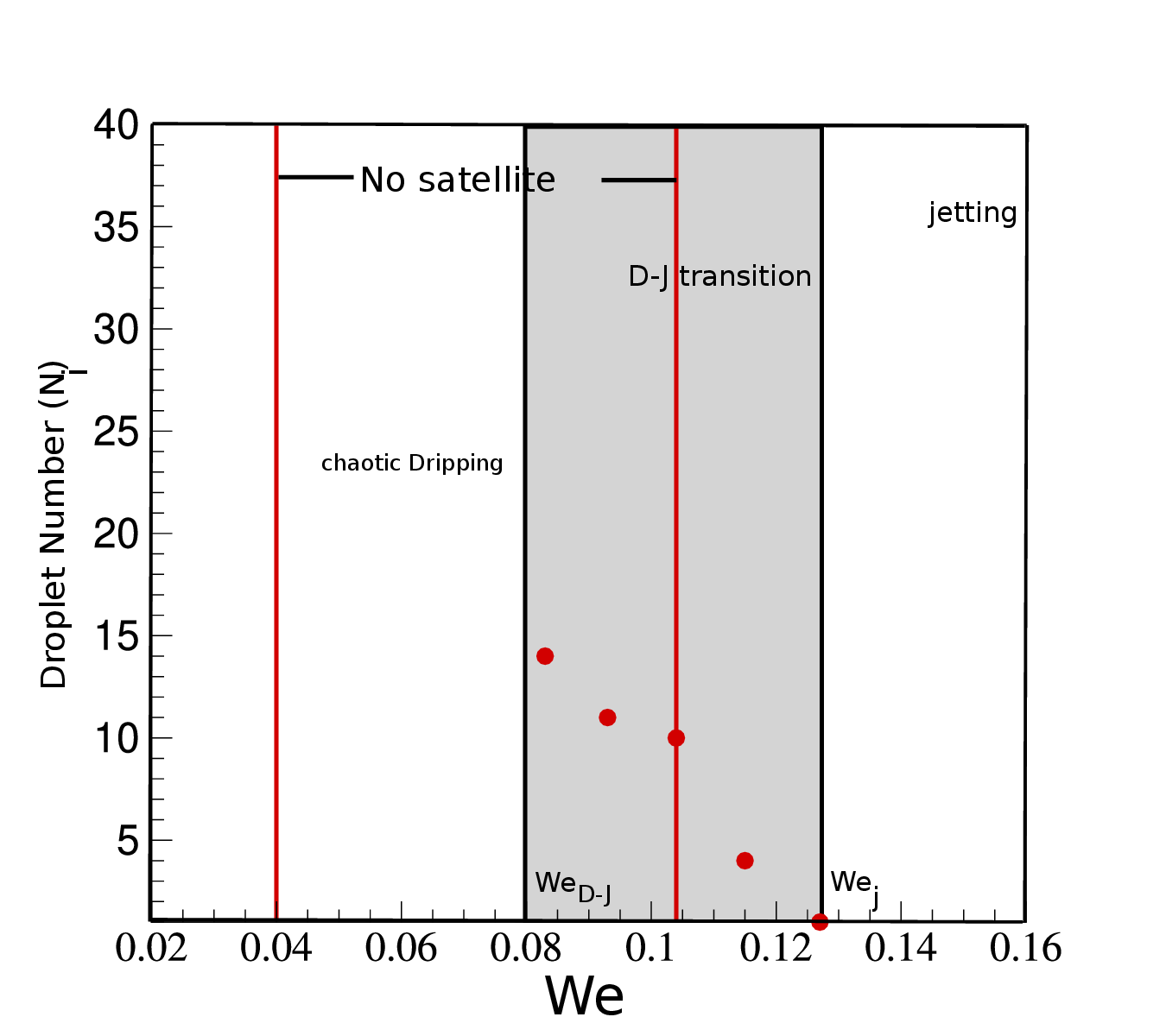} 
\caption{\label{fig:Transient_regime_number_droplets} Phase plot between number of drops before jetting begins as $We$ is increasing for $ Ka = 0.084,  Bo= 1.1$ obtained based on our full Navier-stokes equation.}
\end{figure}

 Figure~\ref{bifurcation}a shows the simulation results for variation in the pinchoff length with $We$ (note that the corresponding inlet velocity is $U_m = \sqrt{We}$). The black dots indicate a forward sweep where the inlet velocity is progressively increased to obtain the transition from dripping regime (left of the shaded region in the figure) to jetting regime (right of the shaded region in the figure). The shaded region marks the hysteresis observed in the quasi-steady state. For the reverse sweep, where the inlet velocity is progressively reduced, we observe that the threshold $We$ for dripping to jetting transition is higher than that for the jetting to dripping transition. The inset shows results from the experiments of Subramani et al.\cite{subramani2006simplicity}. The simulations and experiments for the forward sweep are in good agreement. Figure~\ref{bifurcation}b shows the corresponding droplet volume for different velocities during forward and reverse sweeps. During the reverse sweep  from jetting to dripping, pinchoff length decreases to  $L_b/R_n=10$, and the droplet volume increases to $V_b=15$ at $We = 0.116$ ($U_m=0.34$) whereas for the forward sweep the threshold for dripping to jetting is  $We = 0.336$ ($U_m = 0.58$). As expected, droplet size in the dripping regime is much larger compared to that in the jetting regime (about $15-20$ times). Agreement between the experiments of Subramani et al.\cite{subramani2006simplicity} and simulations for droplet size is also good except for a few points near the transition in the chaotic dripping regime. Away from the transition regime, in both the dripping and jetting regimes the agreement between simulations and experiments is good. These findings are similar to those reported by Rubio et al.~\cite{rubio2018dripping} and Umemura et al.~\cite{umemura2011self} from their experiments. However, as noted earlier, dripping to jetting transition in the steady-state behavior does not reflect the formation of droplets in the dripping regime before the steady jetting regime is achieved for a given inlet velocity $U_m$ ($We$).
Figure \ref{fig:Transient_regime_number_droplets} shows the number of droplet formations before steady state jetting regime is achieved for a range of $We$. Although steady-state critical $We_{D-J}$ is $\sim 0.08$, direct transition from dripping to jetting without any droplet formation in the dripping mode is achieved only at $We \sim  0.135$. For $We = 0.08$, around 15 droplet formations occur in the dripping regime and for $We = 0.105$  jetting is achieved after 10 droplet formations. Thus, in order to control the droplet formation from a nozzle, it is important to know the actual critical $We$ for which no droplet formation occurs in the dripping regime and a jet formation is achieved immediately.\\

\begin{figure}[ht!]
\centering
 \includegraphics[scale=0.7]{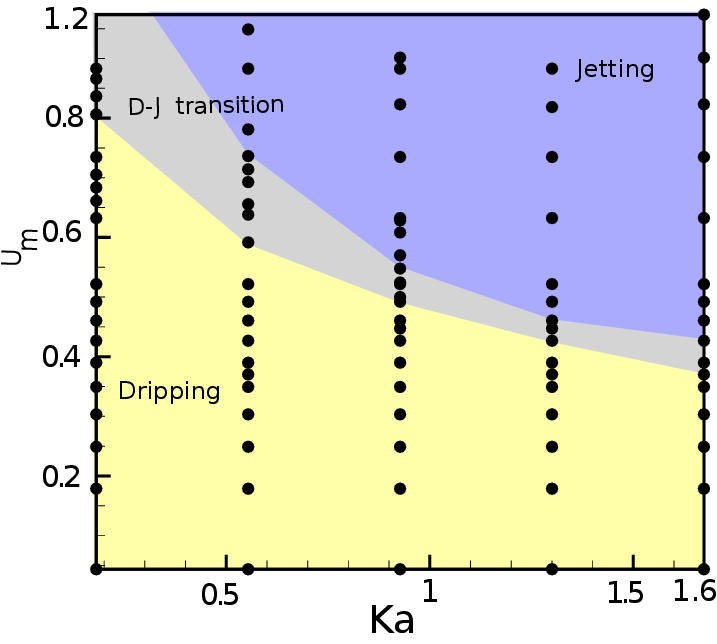} 
\caption{\label{fig:fluid_properties_phase_plot}  Effect of fluid properties on the transition from dripping, dripping-jetting and jetting behaviour of liquid jet at $Bo=0.067$ based on 1D simulations. As the Kapitza number ($Ka$) increases, the transition to dripping-jetting transition has accelerated to 0.48 at $Ka=1.6$. Below 0.48, the breakup happens near the nozzle, and beyond 0.48 to 0.55, the temporal transition from the dripping-jetting transition is observed. The range of Weber number in which this transition occurs has reduced with an increase in $Ka$.}
\end{figure}  

 \begin{figure}
\centering
 \includegraphics[scale=0.85]{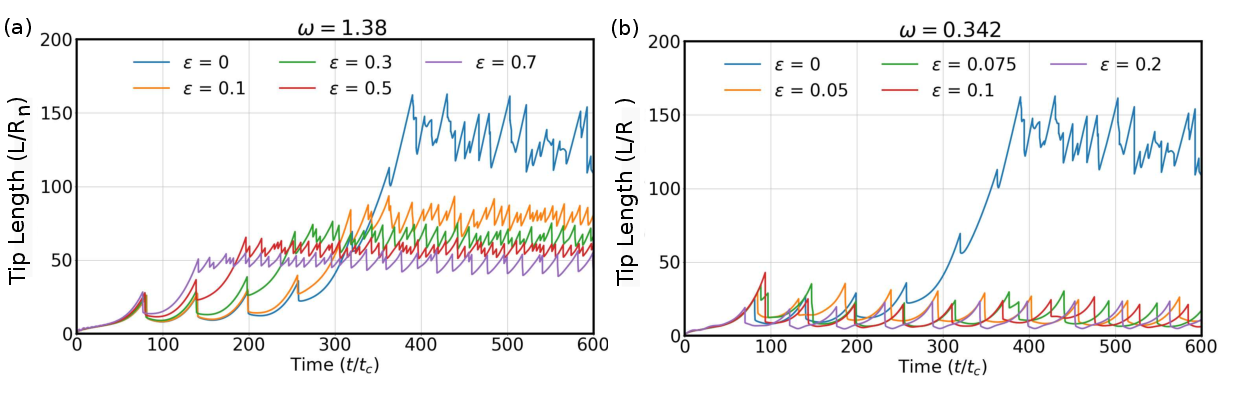} 
\caption{\label{fig:1D_time_length} Influence of velocity perturbation on the temporal evolution and transition of the regime (a) $\omega=0.342$ (b) $\omega=1.38$ for $Ka=1.67$, $We=0.151$ and $Bo=0.067$. }
\end{figure}
\begin{figure}[ht!]
\centering
 \includegraphics[scale=0.881]{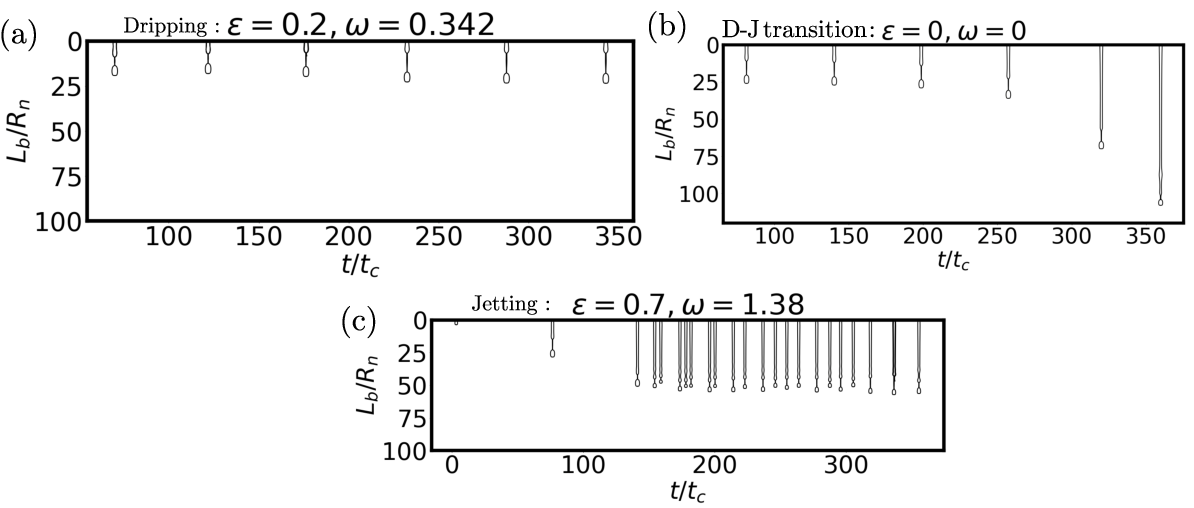} 
\caption{\label{fig:perturbation_profiles} Effect of inlet perturbation on the regime transition is shown. (a) shows dripping regime, (b) D-J transition regime of uncontrolled jet behavior, (C) Jetting regime  at $Bo=0.067$.}
\end{figure}

Figure~\ref{fig:fluid_properties_phase_plot} shows the phase plot showing dripping and jetting regimes as a function of the Weber number ($We$) and Kapitza number ($Ka$) for $Bo=0.067$. At low $Ka$, that is at low viscosities, transition from dripping to jetting occurs at a higher $We$ and a thicker dripping-jetting (D-J) transition region is observed where jetting is achieved after several droplet formations have occurred in the dripping mode of breakup  (for $Ka = 0.181$, $We = 0.64$). The lower boundary can be considered as the dripping-jetting transition boundary if steady-state jetting is considered. With an increase in $Ka$, the critical Weber number (and the corresponding inlet velocity $U_m$) decreases. The region of $We$ over which dripping mode is observed before jetting ensues in the steady state also decreases and for very viscous liquids dripping to jetting transition is observed directly without going through any intermediary regime. In the following, we explore the effect of perturbation in the inlet velocity on the dripping to jetting regime transition. In particular, we study if the dripping-jetting regime can be bypassed by using perturbations of a certain frequency and amplitude.


 In many industrial applications, such as ink jet printing and 3D printing, it is desired to control the size and droplet formation rate. In order to achieve that, it has been proposed that a controlled perturbation be added at the nozzle. This method was applied to generate monodispersed droplet size distributions by \cite{moallemi2016breakup, shukla2020frequency, mcilroy2019effects}. Isha et. al., ~\cite{shukla2020frequency} showed that for an optimal forcing fixed-sized drops can be generated by producing the most unstable jet. Using the shortest breakup length, optimal forcing was identified. These findings are crucial for applications requiring controlled droplet sizes, whether in the millimetre size equal to the nozzle size or in the micron size in the jetting regime, achievable through combinations of $(\epsilon, \omega)$. Here we explore the role of perturbation at the inlet in determining the mode of breakup. We perform simulations using the formulation discussed above modifying the inlet condition to a velocity perturbation:

 \begin{equation}
 a(0, t) = 1 \quad \text{and} \quad u(0, t) = \sqrt{We} + \epsilon \sin(\omega t)
     \label{eq:vel_pert_inlet}
 \end{equation}
 Figure \ref{fig:1D_time_length}a and b show the variation in the length of the jet as a function of time $t/t_c$ for two frequencies $\omega = 1.38$ and $\omega = 3.342$ for different amplitudes $\epsilon$. Pinchoff events are indicated by sharp kinks in the curves. The curve for $\epsilon = 0$ corresponds to the unperturbed condition for the jetting regime ($Ka = 1.67$, $U_m = 0.388$, $We = 0.151$ and $Bo = 0.067$). For a frequency $\omega = 1.38$ (Figure \ref{fig:1D_time_length}a), at a small amplitude of $\epsilon = 0.1$ a reduction in the tip length (length of the contiguous liquid column attached to the nozzle) is observed and the corresponding increase in the number of kinks indicates an increase in the rate of formation of droplets. With an increase in the perturbation amplitude from $\epsilon = 0.1 - 0.7$, the mean tip length progressively decreases and frequency of droplet formation increases. However, the tip length corresponds to the jetting regime only for all values of $\epsilon$. For a higher frequency $\omega = 3.342$, even a small perturbation results in a substantial decrease in the tip length indicating a transition to the dripping regime. The time interval for droplet formation is $\sim 50 t_c$, similar to that observed in the dripping regime. 

Interfacial profiles for $\epsilon = 0.2, \omega = 0.342$ are shown in the figure~\ref{fig:perturbation_profiles}a. The corresponding unperturbed interfacial profiles are shown in figure~\ref{fig:perturbation_profiles}b. The unperturbed interfacial profiles show a progressive increase in the tip length with time indicating a temporal dripping-jetting transition where jetting is achieved in the steady state.  However, if a perturbation is applied with $\epsilon = 0.2, \omega = 0.342$, the jetting regime is avoided and droplet formation occurs in the dripping regime. On the otherhand, if $\epsilon = 0.7, \omega = 1.38$ is used, an earlier transition to the jetting regime is achieved. The frequencies  $\omega = 1.38$ and $\omega = 0.342$ were arbitrarily selected with some cues from the formation frequency of droplets in the dripping regime for the given $Ka$, $Bo$ and $We$.  \\
  

\begin{figure}[ht!]
\centering
 \includegraphics[scale=0.6]{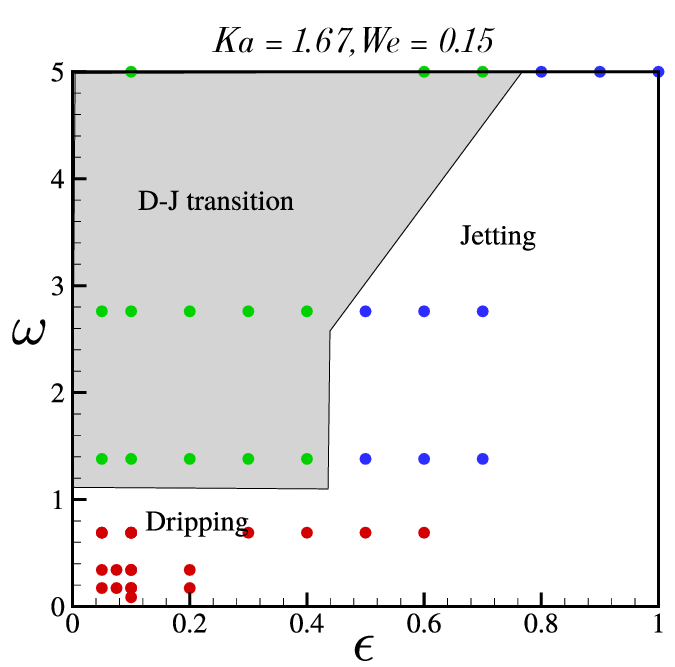} 
\caption{\label{fig:controlling_liquid_jet}  Influence of inlet perturbation parameters ($\epsilon$, $\omega$) on the regime transition at $Bo=0.067$.}
\end{figure} 
Figure~\ref{fig:controlling_liquid_jet}, with $Ka = 1.67$, We = 0.15 ($U_m = 0.39$) , and $Bo = 0.067$ , shows the phase plot for dripping and jetting regime along with dripping-jetting regime where initially dripping is observed and subsequently in the steady state jetting regime occurs. For  $\omega < 1$ dripping regime is observed for all values of $\epsilon$ (in the range 0 to 1). For higher $\omega$, jetting is observed for higher values of $\epsilon$ ($\epsilon > 0.4$). A dripping to jetting transition regime is observed for  $\omega \geq 1$ in the shaded region. 

\section{Conclusion}
\label{Sec:Conclusions}
In this study, we showed that near the threshold value of the $We$ (inlet velocity $U_m$), the mode of droplet formation at a faucet transitions from dripping to jetting mode only after several droplets have formed in the dripping regime. The number of droplets that form before transition occurs is influenced by the fluid properties $Ka$ and nozzle diameter $Bo$. For higher $Ka$, we observe that the transition to jetting can occur at a lower Weber number and without formation of any droplets in the dripping regime. 

We showed that the mechanism of the transition from dripping to jetting involves a progressive elongation of the liquid column formed after each droplet breakup. This elongation of the liquid column occurs through tip retraction and delay in the pinchoff, here termed as the escape pinchoff mechanism, similar to that discussed in \cite{hoepffner2013recoil} during recoil of a liquid column. The above mechanism alters the droplet shape, resulting in prolate and oblate shape oscillations of the droplet. This may cause droplets to have larger volumes during the transition regime. Numerical simulations show that for a fixed Bond number, the transition region reduces and jetting regime is achieved without initial transitions (droplet formations in the dripping regime) for higher values of the inlet velocity $U_m$ (i.e. $We$). Further, we show that the dripping-jetting transition regimes can be significantly altered by introducing perturbations in the inlet velocity and desired droplet size distribution can be achieved by using perturbations of a certain frequency and amplitude.

\section*{Acknowledgments}
GT acknowledges the FIST grant of the Department of Science and Technology India, for the high speed camera Photron s9000 series used in this study.
 
\appendix
 
\section{Gerris simulation validation}
\label{appA}
\begin{figure} 
\centering
\captionsetup{width=1\linewidth}
\captionsetup{justification=justified}  \includegraphics[width=1\textwidth]{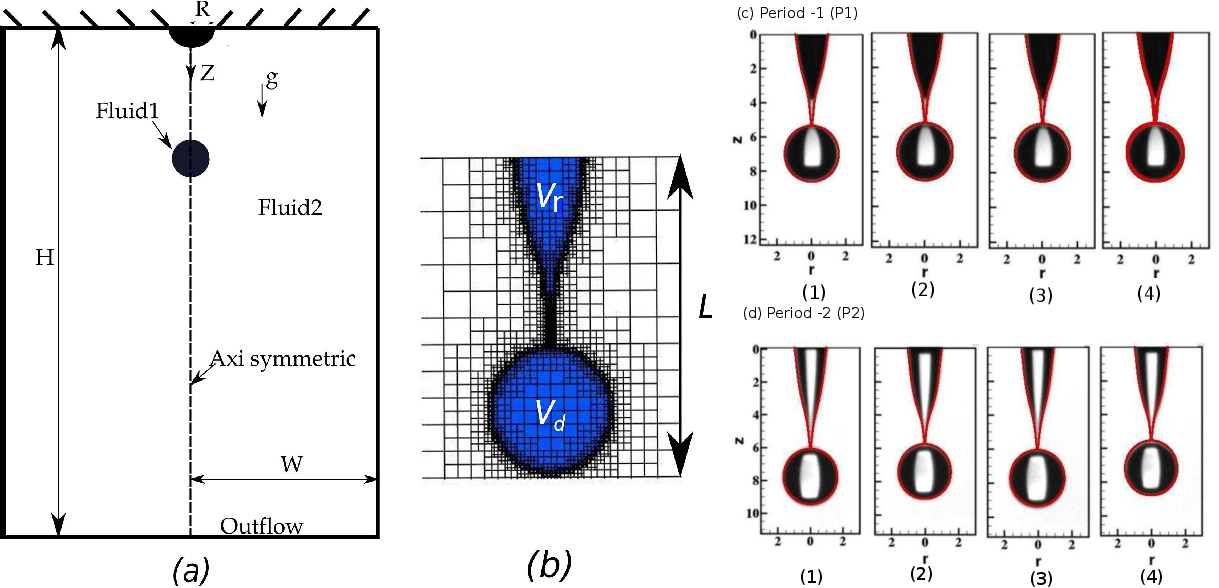}
\caption{  Schematic showing fluid droplet from a nozzle of diameter $R$ is shown. $(d)$ Computational domain of length $H$ and width $W$ with the nozzle placed at the axisymmetry boundary. A liquid droplet pinchoff from the nozzle is depicted with Fluid1 is injected into the computational domain from above. A coordinate frame ($r-z$ plane) is attached at the axis of symmetry with $z$-coordinate pointing in the direction of the gravitational acceleration ($\mathbf{g}$). Ambient fluid is air (with density $\rho_g = 1kg/m^3$) and is assumed to be quiescent. Outflow boundary conditions are imposed at the bottom end. (b) Droplet evolved near the pinchoff with AMR grid showing with length L and  left over liquid volume $V_r$ with a droplet volume of $V_d$. The interface profiles of simulations compared with the experimental results of Subramani et al. \cite{subramani2006simplicity} for both P1 and P2 regimes in figure c, d.
  }
\label{fig:Gerris_Domain}
\end{figure}

We perform numerical simulations using volume-of-fluid method in the two-phase flow open source code Gerris (\cite{popinet2003gerris,popinet2009accurate,tomar2010multiscale}). Computational domain chosen for the simulations is shown in Figure \ref{fig:Gerris_Domain}a. We perform axisymmetric simulations and thus only half of the domain, defined by length $L$ and width $W$, as shown in the figure is employed for performing the simulations. The gravitational acceleration is in the $z$ direction (downward) and the ambient fluid is air.
 
 Gerris flow solver employed in the present study is based on a one-fluid formulation. Both the liquid and the ambient air are considered incompressible. The Navier-Stokes equations for momentum conservation are scaled with the nozzle diameter chosen as the characteristic length ($D_n$) and the inlet velocity ($v$) chosen as the velocity scale. The Navier-Stokes equation is modified to include the surface tension force term (with surface tension coefficient $\sigma$) acting on the interface (embedded in the Eulerian grid) and is expressed as a volumetric force following the continuum surface force (CSF) model by employing a surface Dirac delta function ($\mathbf{\delta_{s}}$) \cite{brackbill1992continuum}. The resulting non-dimensionalized equations for incompressibility and momentum conservation are respectively given as, 
    \begin{equation}
      \nabla \cdot \mathbf{u}=0 ~~\mbox{and}
  \label{continuity}
  \end{equation}

    \begin{equation}
      \rho \left[ \frac {\partial \mathbf{u}}{\partial t}+\left( \mathbf{u}\cdot \nabla \right)  \mathbf{u}\right] =-\nabla P+\left( \frac {\kappa}{We}\right)\mathbf{n}\delta_{s} -\rho \left( \frac {Bo}{We}\right) \\+\left( \frac {1}{Re}\right) \nabla \cdot \left[ \mu \left( \nabla \mathbf{u}+\nabla \mathbf{u}^{T}\right) \right]
       \label{NS}
\end{equation}
where $\kappa$ and $\mathbf{n}$ are the curvature and the unit normal at the interface, respectively. The non-dimensional numbers governing the problem are, Weber number $We = {\rho_l v^2 D_n}/{\sigma}$, Reynolds number $Re = {\rho_l v D_n}/{\mu_l}$ and Bond number $Bo = {\rho_l g D_n^2}/{\sigma}$. The density ($\rho$) and viscosity ($\mu$) in the above equation (Eq.\ref{NS}) are scaled by the liquid properties and are a function of the local liquid fraction (ratio of the liquid volume in a given Eulerian grid cell to the volume of the grid cell), $\alpha$,    
\begin{equation} 
    \rho =\alpha +\left( 1-\alpha \right) (\rho _{a}/\rho _{l})
\end{equation}
\begin{equation}
     \mu =\alpha +\left( 1-\alpha \right) (\mu _{a}/\mu _{l})
\end{equation}

 The density ratios $\rho_a/\rho_l=1.5 \times 10^{-3}$ and $\mu_a/\mu_l=3.7 \times 10^{-4}$ are fixed based on the liquid properties given in the experimental study by \cite{subramani2006simplicity}. The interface evolution is captured by solving the advection equation for volume fraction $\alpha$ using geometric volume of fluid method,
\begin{equation}
         \frac {\partial \alpha }{\partial t}+ \boldsymbol{u}\cdot\nabla\alpha =0
\end{equation}

 A uniform flow with velocity $v$ from the nozzle region ($D_n$) on the top surface is imposed as the inflow condition. A  no-slip boundary condition is imposed on the fiber surface. At the bottom surface of the domain outflow boundary condition is imposed, that is, Dirichlet condition on the pressure ($P=0$), and Neumann condition on the axial and radial velocities, given by $\partial_zU = 0$ and $\partial_zV =0$, respectively. Axi-symmetric boundary conditions are imposed at the $r=0$ surface and slip boundary condition is imposed on the $r = W$ surface. The length of the computational domain is chosen to be $L = 100 mm$ and $W = 8mm$. The adaptive mesh refinement is employed with the interface being resolved using a fine mesh size of $\delta=32.8 \mu m$, whereas a coarser mesh ($\delta=262.5 \mu m$) is employed in the gas region away from the interface. 
A study of the influence of the mesh was conducted to rule out the effect of spatial resolution on the film evolution. Figure~\ref{fig:Gerris_Domain}b  shows the film interface profiles at pinchoff conditions with AMR grid with Levels 5 and 8 (maximum refinement regions). All simulation results presented in this study correspond to Level 8. The final simulations results are compared with the experimental results of \cite{subramani2006simplicity} in figure \ref{fig:Gerris_Domain}c,d where the simulations profiles are overlapped onto the experimental pinchoff conditions. The simulation results accurately compare the period 1 and period 2 regimes. 
\section{Kernel Density Estimation of Droplet Size Distribution}
\label{appB}
 Kernel Desnsity Estimation (KDE) is used to visualize the droplet diameter distribution more smoothly on a histogram as shown in the figure \ref{fig:Droplet_size_distribution}. This method estimates the probability density function by  placing kernal (Gaussian curve) at each data poin and summing them.  The bandwidth which controls the smoothness of the curve, is selected using $ h = \sigma n^{-1/5}$ rule. Where h is the bandwith, $\sigma$  is the standard deviation of the data, and n is the number of droplet diameters.  
\bibliographystyle{apsrev4-1}
\bibliography{bibfile}  

\end{document}